\documentstyle[pre,aps]{revtex}
\input{epsf}

\begin{document}
\title{Long Range Correlations in the Disordered Phase of a Simple Three State
Lattice Gas}
\author{G. Korniss, B. Schmittmann and R. K. P. Zia}
\address{Center for Stochastic Processes in Science and Engineering \\
and \\
Department of Physics, Virginia Polytechnic Institute and State \\
University \\
Blacksburg, Virginia 24061-0435, USA}
\date{June 10, 1996}
\maketitle

\begin{abstract}
We investigate the dynamics of a three-state stochastic lattice gas,
consisting of holes and two oppositely ``charged" species of particles,
under the influence of an ``electric" field, at zero total charge.
Interacting only through an excluded volume constraint, particles can hop to
nearest neighbour empty sites. With increasing density and drive, the system
orders into a charge-segregated state. Using a combination of Langevin
equations and Monte Carlo simulations, we study the steady-state structure
factors in the {\em disordered} phase where homogeneous configurations are
stable against small harmonic perturbations. They show a discontinuity
singularity at the origin which in real space leads to an intricate
crossover between power laws of different kinds.
 \\

\end{abstract}

\vfill

For a system with short range microscopics, placed in thermal equilibrium,
long range spatial correlations are absent in general. Their presence is a
typical signal that the system is at a singular point. On the other hand,
for systems in non-equilibrium steady states, long range correlations are
often observed\cite{liquids}. With conserved dynamics they generically arise
if the system has strong spatial anisotropies and lacks detailed balance. A
good example is the driven Ising lattice gas \cite{BS_RKPZ,KLS}. The role of
the drive (``external field'') is to bias hopping rates along a particular
direction on the lattice. In addition to many other unexpected features such
as non-Hamiltonian fixed points controlling the critical behavior\cite{JS_LC}%
, the system exhibits long range spatial correlations {\em at all
temperatures }above criticality, as a result of the breakdown of the
fluctuation-dissipation theorem (FDT). In momentum space, this appears as a
discontinuity singularity of the structure factor at the origin \cite{ZHSL}.

One can generalize the model of ref \cite{KLS} in a fashion similar to the
one leading to the Ising model to spin-1 \cite{BEG} or Potts \cite{Potts}
models, by considering more than one species of particles. To start with, we
have two species of particles (referred as $+$'s and $-$'s) driven in
opposite directions and holes (empty sites). To keep the model simple,
however, we may neglect the usual Ising nearest neighbor interaction and
keep ``only'' the excluded volume constraint and the bias. This
multi-species model, in both one and two dimensions, has been thoroughly
investigated \cite{SHZ,VZS,FG,KSZ,traffic,obc}. In the simplest scenario
charge ($+-$) exchange is not allowed. Monte Carlo simulations \cite{SHZ} in
two dimensions and mean-field studies \cite{VZS,FG} show that there is a
transition, controlled by particle density and drive, from a spatially
homogeneous (disordered) phase to a charge segregated one, where the
excluded volume constraint leads to the mutual blocking of particles. If we
soften the excluded volume constraint -- by allowing exchange between
nearest neighbor, oppositely charged particles on a much slower time scale
than the particle hole exchange -- the above transition still survives \cite
{KSZ}. In the {\em disordered} phase, these two versions behave in much the
same way. Thus, for simplicity, we will study only the case {\em without}
charge exchange in this paper and focus on the spatial correlations and
structure factors.

We consider a two dimensional fully periodic lattice with $L_{\perp }\times
L_{\parallel }$ sites, each of which can be empty or occupied by a single
particle. Since we have two species we need two occupation variables $n_{%
{\bf x}}^{+}$ and $n_{{\bf x}}^{-}$, with $n$ being 0 or 1, depending on
whether a positive or negative particle is present at site ${\bf x}$.
Although we refer to these particles as ``charged'', they do not interact
via the Coulomb potential. Instead, they respond to an external, uniform
electric field $E$, directed along the $+x_{\parallel }$-axis. We restrict
ourselves to zero total charge, i.e. $\sum_{{\bf x}}[n_{{\bf x}}^{+}-n_{{\bf %
x}}^{-}]=0$. In the absence of the drive the dynamics does not distinguish
between the different species: both types hop randomly to nearest neighbor
empty sites, with a rate $\Gamma $. The electric field introduces a bias
into these rates in such a way that jumps {\em against} the force will be
exponentially suppressed. During one Monte Carlo step $2L_{\perp
}L_{\parallel }$ nearest neighbor bonds are picked randomly. If a
particle-hole pair is encountered, an exchange takes place with probability $%
W=\Gamma \min \{1,\exp (qE\,\delta x_{\parallel })\}$, where $q=\pm 1$ is
the charge of the particle and $\delta x_{\parallel }=\pm 1$ is the change
of the $x_{\parallel }$ coordinate of the particle due to the jump.

For our simulations, we set $\Gamma=1$.
Using $30\times 30$ and $60\times 60$ lattices, we initialized the system
with random configurations of various particle densities and carried out
runs ranging from $2.5$ to $5\times 10^5$ MCS. After allowing 62500 MCS for
the system to settle into a steady state, we measured the Fourier transform
of $n_{{\bf x}}^{\pm }$ every $125$ MCS. The structure factors are then
constructed from averages of these measurements, according to 
\begin{equation}
S^{\alpha \beta }({\bf k})=\frac 1{L_{\perp }L_{\parallel }}\sum_{{\bf x},%
{\bf y}}e^{-i\,{\bf k\cdot }({\bf x}-{\bf y})}\langle n_{{\bf x}}^\alpha n_{%
{\bf y}}^\beta \rangle \;,  \label{simul_def}
\end{equation}
where $\alpha ,\beta =+,-$ and ${\bf k}=(\frac{2\pi m_{\perp }}{L_{\perp }},%
\frac{2\pi m_{\parallel }}{L_{\parallel }})\neq {\bf 0}${\bf \ }. Note that $%
S^{\alpha \beta }$ is the Fourier transform of the usual {\em equal-time }%
correlation function $G^{\alpha \beta }({\bf x})\equiv \langle n_{{\bf x}%
}^\alpha n_{_{{\bf 0}}}^\beta \rangle -\langle n_{{\bf x}}^\alpha
\rangle \langle n_{_{{\bf 0}}}^\beta \rangle $. Thus, if $G$
is even in ${\bf x}$, $S$ will be real. Equivalently, an imaginary part of $S
$ corresponds to the part of $G$ which is odd in ${\bf x}$. Due to charge
symmetry, we expect $G^{++}=G^{--}$. Clearly, both must be even in ${\bf x}$%
, so that the associated $S$'s are real. On other hand, due to the drive, we
have

\begin{equation}
G^{+-}({\bf x,}E)=G^{+-}(x_{_{\perp }}{\bf ,-}x_{\parallel }\text{ }{\bf ,-}%
E)  \label{g+-}
\end{equation}
so that $S^{+-}$ may have an imaginary part (which must be odd in $E$).
Finally, $G^{-+}({\bf x})=G^{+-}(-{\bf x})$ is just a mathematical identity.

\begin{figure}[tbp]
\hspace*{2cm}
\epsfxsize=12cm
\epsfysize=12cm
\epsfbox{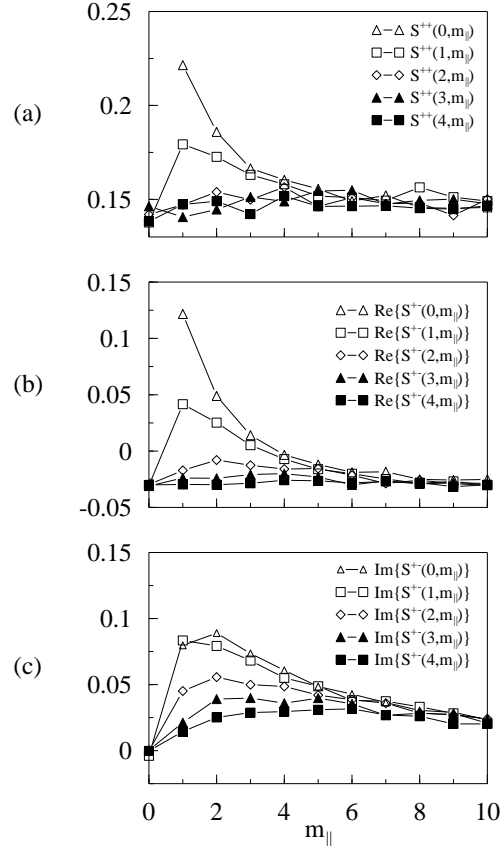}
\caption{Steady state structure factors (a) $S^{++}({\bf k})$ , (b) $
\mbox{Re}\{S^{+-}({\bf k})\}$ , (c) $\mbox{Im}\{S^{+-}({\bf k})\}$ for a $%
60\times 60$ system at $E=0.471$ and $\bar{\varrho}=0.175$. Structure
factors are plotted against the integer $m_{\parallel}=\frac{k_{\parallel}
L_{\parallel}}{2\pi}$, while $m_{\perp}=\frac{k_{\perp} L_{\perp}}{2\pi}$ is
taken as a parameter. }
\label{fig1}
\end{figure}

In Fig. \ref{fig1}, we present the results for the three independent $S$'s
found in the larger system. Before discussing the data in detail, we will
first present the theoretical framework within which they can be understood.
In particular, we will illustrate the emergence of discontinuity
singularities in the structure factors at ${\bf k=0}$, and their
consequences for long-range correlations in real space. This will then be
followed by a comparison between our theoretical predictions and the
simulation results.

Since our interest lies in the behavior at large distances (or small ${\bf k}
$), the simplest approach relies on continuum field theory. Starting from
the dynamics at the microscopic level, specified by the above rates and the
associated Master equation, there are several ways to arrive at a
coarse-grained description, in terms of equations of motions for the local $t
$-dependent densities. The most systematic one is to perform an $\Omega $%
-expansion \cite{VK}, where one splits the particle densities describing a
block b (size $\Omega $), around point ${\bf {x}}$, into a macroscopic part (%
$\varrho ^{\pm }$) and a fluctuating one ($\chi ^{\pm }$) : 
\begin{equation}
\frac 1\Omega \sum_{{\bf x}^{^{\prime }}\,\epsilon \, \mbox{b}({\bf x}%
)}n_{{\bf x}^{^{\prime }}}^{\pm }=\varrho ^{\pm }({\bf x},t)+\Omega
^{-1/2}\chi ^{\pm }({\bf x},t)\;.  \label{omega}
\end{equation}
For generality, we consider the d-dimensional case when ${\bf x}_{\parallel }
$ is directed along the electric field and ${\bf x}_{\perp }$ is in the d-1
dimensional subspace, perpendicular to the field. After taking the continuum
limit the result for the macroscopic part is a set of mean-field equations
of motions for the slowly varying local densities: 
\begin{equation}
\partial _t\varrho ^{\pm }=-\mbox{\boldmath $\nabla$}{\bf \Gamma }\{\varrho
^{\pm }\stackrel{\leftrightarrow }{\mbox{\boldmath $\nabla$}}(1-\varrho
^{+}-\varrho ^{-})\pm \varepsilon \hat{{\bf x}}_{\parallel }\varrho ^{\pm
}(1-\varrho ^{+}-\varrho ^{-})\};\,  \label{meaf}
\end{equation}
where 
\begin{equation}
{\bf \Gamma }=\left( \matrix{ {\bf \Gamma}_{\perp} & {\bf 0} \cr {\bf 0} &
\Gamma_{\parallel} }\right)   \label{diffmatr}
\end{equation}
is the diffusion-matrix. ${\bf \Gamma }_{\perp }$ is diagonal and isotropic
in the d-1 dimensional subspace, thus characterized by a number $\Gamma
_{\perp }$. $\stackrel{\leftrightarrow }{\mbox{\boldmath $\nabla$}}$ is the
asymmetric gradient operator, namely for any two functions $f$ and $g$, $f%
\stackrel{\leftrightarrow }{\mbox{\boldmath $\nabla$}}g=f%
\mbox{\boldmath
$\nabla$}g-g\mbox{\boldmath $\nabla$}f$. $\varepsilon =2\tanh (E/2)$ is the
coarse-grained bias and $\hat{{\bf x}}_{\parallel }$ is the unit vector
along the $x_{\parallel }$ direction.

Since (\ref{meaf}) is a continuity equation, it trivially admits $t$%
-independent solutions which are homogeneous in space: $\varrho ^{\pm }({\bf %
x},t)=\bar{\varrho}$ . For simplicity, and to ease comparison with
simulation data, we have chosen equal densities for both species. At the
mean-field level, this solution describes the steady state. The correlations
of interest are then associated with $\chi ^{\pm }$, the scaled deviations (%
\ref{omega}) from the mean-field steady state. 

Returning to the $\Omega $-expansion, the result for the fluctuating part,
at lowest order, is a Fokker-Planck equation. For our purposes, the
equivalent Langevin equation is more transparent. At this order, its
deterministic part is linear and the (conserved) noise is Gaussian. Focusing
on the fluctuations about the homogeneous phase, the result is: 
\begin{equation}
\partial _t\left( \matrix{ \chi^{+}({\bf x},t) \cr \chi^{-}({\bf x},t) }%
\right) ={\cal L}(\mbox{\boldmath $\nabla$})\left( \matrix{ \chi^{+}({\bf
x},t) \cr \chi^{-}({\bf x},t) }\right) -\left( 
\matrix{ \mbox{\boldmath
$\nabla\eta $}^{+}({\bf x},t) \cr \mbox{\boldmath $\nabla\eta $}^{-}({\bf x},t) }%
\right) \;,  \label{rsl}
\end{equation}
where 
\begin{eqnarray}
{\cal L}(\mbox{\boldmath $\nabla$}) &=&\left( 
\matrix{ {\cal L}^{++}(\mbox{\boldmath $\nabla$}) & {\cal L}^{+-}(\mbox{\boldmath
$\nabla$}) \cr {\cal L}^{-+}(\mbox{\boldmath $\nabla$}) & {\cal L}^{--}(\mbox{\boldmath $\nabla$}) }%
\right) =  \nonumber \\
&&  \nonumber \\
&&\left( \matrix{ (1-\bar{\varrho})\mbox{\boldmath $\nabla\Gamma\nabla$} -
(1-3\bar{\varrho}) \varepsilon\Gamma_{\parallel}\partial_{\parallel} &
\bar{\varrho}\mbox{\boldmath $\nabla\Gamma\nabla$} + \bar{\varrho}
\varepsilon\Gamma_{\parallel}\partial_{\parallel} \cr
\bar{\varrho}\mbox{\boldmath $\nabla\Gamma\nabla$} - \bar{\varrho}
\varepsilon\Gamma_{\parallel}\partial_{\parallel} &
(1-\bar{\varrho})\mbox{\boldmath $\nabla\Gamma\nabla$} + (1-3\bar{\varrho})
\varepsilon\Gamma_{\parallel}\partial_{\parallel} }\right) \;.
\label{lin_def}
\end{eqnarray}
Here, $\mbox{\boldmath $\eta $}^{\pm }({\bf x},t)$ are Gaussian, white noise
terms, satisfying: 
\begin{eqnarray}
\langle \eta _i^\alpha ({\bf x},t)\rangle =0 &\;,\;\;\;\;\;&\langle \eta
_i^\alpha ({\bf x},t)\eta _j^\beta ({\bf x}^{\prime },t^{\prime })\rangle
=2\delta ^{\alpha \beta }\sigma _{ij}\delta ({\bf x}-{\bf x}^{\prime
})\delta (t-t^{\prime })  \label{rsnoise} \\
\alpha ,\beta =+,- &;&i,j=1,2,\ldots d\;.  \nonumber
\end{eqnarray}
$(\sigma _{ij})=\mbox{\boldmath $\sigma$}$ is the noise matrix: 
\begin{equation}
\mbox{\boldmath $\sigma$}=\left( \matrix{ \mbox{\boldmath $\sigma$} _{\perp}
& {\bf 0} \cr {\bf 0} & \sigma_{\parallel} }\right) \;.  \label{noisematr}
\end{equation}
Similar to ${\bf \Gamma }_{\perp }$, $\mbox{\boldmath $\sigma$}_{\perp }$ is
diagonal and isotropic in the d-1 dimensional subspace, characterized by a
number $\sigma _{\perp }$. In equilibrium systems, the FDT guarantees $%
\mbox{\boldmath $\sigma$}\propto {\bf \Gamma }$. In particular, in the
absence of the drive, we would have $\mbox{\boldmath $\sigma$}=\bar{\rho}(1-2%
\bar{\rho}){\bf \Gamma }$ here. However, when driven, the proportionality is
not expected to hold, in that the diffusion and noise matrices would be
renormalized differently by the drive $\varepsilon $. This is certainly the
situation in the driven single species case\cite{JS_LC}. Finally, we point
out that $\mbox{\boldmath $\eta $}^{+}$ and $\mbox{\boldmath $\eta $}^{-}$
are uncorrelated, due to the fact that charge exchange is not allowed.

To find the correlations and structure factors from eqns. (\ref{rsl}-\ref
{rsnoise}), we introduce the Fourier components for the fluctuations: 
\begin{equation}
\chi ^{\pm }({\bf k},\omega )=\int dtd^d{\bf x}\text{ }\chi ^{\pm }({\bf x}%
,t)\text{ }e^{-i(\omega t+{\bf kx})}\;,  \label{ftrdef}
\end{equation}
and similar ones for the noise, so that 
\begin{eqnarray}
\langle \eta _i^\alpha ({\bf k},\omega )\rangle =0 &\;,\;\;\;\;\;&\langle
\eta _i^\alpha ({\bf k},\omega )\eta _j^\beta ({\bf k}^{\prime },\omega
^{\prime })\rangle =2\delta ^{\alpha \beta }\sigma _{ij}\left[ (2\pi
)^{d+1}\delta ({\bf k}+{\bf k}^{\prime })\delta (\omega +\omega ^{\prime
})\right]  \label{ksnoise} \\
\alpha ,\beta =+,- &;&i,j=1,2,\ldots d\;.  \nonumber
\end{eqnarray}
Then the solution to (\ref{rsl}) is trivial: 
\begin{equation}
\left( \matrix{ \chi^{+}({\bf k},\omega) \cr \chi^{-}({\bf k},\omega) }%
\right) =({\cal L}(i{\bf k})-i\omega )^{-1}\left( 
\matrix{ i{\bf k}\mbox{\boldmath $\eta $}^{+}({\bf k},\omega) \cr 
i{\bf k}\mbox{\boldmath $\eta
$}^{-}({\bf k},\omega) }\right) \;.  \label{inv_ksl}
\end{equation}
Note that, in ${\bf k}$ space, $\left( {\cal L}^{++},{\cal L}^{--}\right) $
and $\left( {\cal L}^{+-},{\cal L}^{-+}\right) $ are complex conjugate pairs.

As expected, $\langle \chi ^{\pm }({\bf k},\omega )\rangle =0$, since these
are the fluctuations about the conserved average densities. Their
correlations are just the {\em dynamic} structure factors, easily obtained
from (\ref{ksnoise}) and (\ref{inv_ksl}). From the definition: 
\begin{equation}
\langle \chi ^\alpha ({\bf k},\omega )\chi ^\beta ({\bf k}^{\prime },\omega
^{\prime })\rangle \equiv S^{\alpha \beta }({\bf k},\omega )\left[ (2\pi
)^{d+1}\delta ({\bf k}+{\bf k}^{\prime })\delta (\omega +\omega ^{\prime
})\right]  \label{dyn_str_def}
\end{equation}
we find the two independent $S$'s: 
\begin{eqnarray}
S^{++}({\bf k},\omega ) &=&\,\frac{2\,\mbox{\boldmath $k \sigma k$}}{\mid
\Lambda ({\bf k},\omega )\mid ^2}\,\{\mid i\omega -{\cal L}^{--}(i{\bf k}%
)\mid ^2+\mid {\cal L}^{+-}\mid ^2\}  \nonumber \\
S^{+-}({\bf k},\omega ) &=&\,\frac{2\,\mbox{\boldmath $k \sigma k$}}{\mid
\Lambda ({\bf k},\omega )\mid ^2}\,\{-2{\cal L}^{--}(i{\bf k}){\cal L}^{+-}(i%
{\bf k})\}  \label{dyn_str}
\end{eqnarray}
where 
\begin{equation}
\Lambda ({\bf k},\omega )=\det ({\cal L}(i{\bf k})-i\omega )\;.  \label{det}
\end{equation}
To find the equal-time correlations, we need the steady state structure
factors, which can be obtained from (\ref{dyn_str}) by an integration over $%
\omega $ . The results are: 
\begin{eqnarray}
S^{++}({\bf k}) &=&\,\frac{\mbox{\boldmath
$k \sigma k$}}{-\frac 12\mbox{Tr}{\cal L}(i{\bf k})}\,\frac{\mid {\cal L}%
^{++}(i{\bf k})\mid ^2}{\det {\cal L}(i{\bf k})}  \nonumber \\
S^{+-}({\bf k}) &=&\,\frac{\mbox{\boldmath $k \sigma k$}}{-\frac 12\mbox{Tr}%
{\cal L}(i{\bf k})}\,\frac{-{\cal L}^{--}(i{\bf k}){\cal L}^{+-}(i{\bf k})}{%
\det {\cal L}(i{\bf k})}\;.  \label{steady_str}
\end{eqnarray}
Note that $-\frac 12\mbox{Tr}{\cal L}(i{\bf k})=(1-\bar{\varrho}){\bf %
k\Gamma k}$ is positive definite, while we need to require $\det {\cal L}(i%
{\bf k})>0$ for all ${\bf k}\neq 0$ to assure that the system is within the
linear stability boundary. Finally , using (\ref{lin_def}), eqs. (\ref
{steady_str}) take the explicit form: 
\begin{eqnarray}
S^{++}({\bf k}) &=&\,\frac{(1-\bar{\varrho})}{(1-2\bar{\varrho})}\,\frac{%
\mbox{\boldmath $k \sigma k$}}{{\bf k\Gamma k}}\;\frac{({\bf k\Gamma k})^2+%
\frac{(1-3\bar{\varrho})^2}{(1-\bar{\varrho})^2}\Gamma _{\parallel
}^2\varepsilon ^2k_{\parallel }^2}{({\bf k\Gamma k})^2+(1-4\bar{\varrho}%
)\Gamma _{\parallel }^2\varepsilon ^2k_{\parallel }^2}  \nonumber \\
\mbox{Re}\{S^{+-}({\bf k})\} &=&-\,\frac{\bar{\varrho}}{(1-2\bar{\varrho})}\,%
\frac{\mbox{\boldmath $k \sigma k$}}{{\bf k\Gamma k}}\;\frac{({\bf k\Gamma k}%
)^2-\frac{(1-3\bar{\varrho})}{(1-\bar{\varrho})}\Gamma _{\parallel
}^2\varepsilon ^2k_{\parallel }^2}{({\bf k\Gamma k})^2+(1-4\bar{\varrho}%
)\Gamma _{\parallel }^2\varepsilon ^2k_{\parallel }^2}  \label{exp_str} \\
\mbox{Im}\{S^{+-}({\bf k})\} &=&\,\frac{2\bar{\varrho}}{(1-\bar{\varrho})}\;%
\frac{\mbox{\boldmath $k \sigma k$}\,\Gamma _{\parallel }\varepsilon
k_{\parallel }}{({\bf k\Gamma k})^2+(1-4\bar{\varrho})\Gamma _{\parallel
}^2\varepsilon ^2k_{\parallel }^2} \; . \nonumber
\end{eqnarray}

A key feature of these structure factors is that, unlike in equilibrium
cases, all are singular at the origin. For both $S^{++}$ and $\mbox{Re}%
\{S^{+-}\} $, this singularity is exhibited as a discontinuity, i.e., $%
\lim_{k_{\parallel }\rightarrow 0}S({\bf 0},k_{\parallel })\neq \lim_{{\bf k}%
_{\perp }\rightarrow {\bf 0}}S({\bf k}_{\perp },0).$ In particular, 
\begin{eqnarray}
\frac{\lim_{k_{\parallel }\rightarrow 0}S^{++}({\bf 0},k_{\parallel })}{%
\lim_{{\bf k}_{\perp }\rightarrow {\bf 0}}S^{++}({\bf k}_{\perp },0)}
&=&\left[ \frac{\sigma _{\parallel }}{\Gamma _{\parallel }}\frac{\Gamma
_{\perp }}{\sigma _{\perp }}\right] \frac{(1-3\bar{\varrho})^2}{(1-\bar{%
\varrho})^2(1-4\bar{\varrho})}  \label{S++_disc}
\end{eqnarray}
and 
\begin{eqnarray}
\frac{\lim_{k_{\parallel }\rightarrow 0}\mbox{Re}\{S^{+-}({\bf 0}%
,k_{\parallel })\}}{\lim_{{\bf k}_{\perp }\rightarrow {\bf 0}}\mbox{Re}%
\{S^{+-}({\bf k}_{\perp },0)\}}=-\ \left[ \frac{\sigma _{\parallel }}{\Gamma
_{\parallel }}\frac{\Gamma _{\perp }}{\sigma _{\perp }}\right] \frac{(1-3%
\bar{\varrho})}{(1-\bar{\varrho})(1-4\bar{\varrho})}  \; .
\label{S+-_disc}
\end{eqnarray}

In general these ratios are not unity. Note that these singularities come
not only from the generic FDT-breaking property, namely $\frac{\sigma
_{\parallel }}{\Gamma _{\parallel }}\neq \frac{\sigma _{\perp }}{\Gamma
_{\perp }}$, but also from the specifics of this particular driven system.
This second factor is a monotonically increasing function of $\bar{\varrho}$
reaching $\infty $ at $\bar{\varrho}$ =1/4. This divergence is related to
the instability of the homogeneous phase, as we will see below.

On the other hand, though$\ \mbox{Im}\{S^{+-}({\bf k})\}$ vanishes for ${\bf %
k}\rightarrow {\bf 0}$ in any direction, discontinuities are present in
higher derivatives. In configuration space, these singularities translate
into power law decays of the equal-time correlation functions $\langle \chi
^\alpha ({\bf x}^{\prime }+{\bf x},t)\chi ^\beta ({\bf x}^{\prime
},t)\rangle $, which is, thanks to translational invariance, independent of $%
{\bf x}^{\prime }$. More precisely, it is just

\begin{equation}
G^{\alpha \beta }({\bf x})=\int \frac{d^d{\bf k}}{(2\pi )^d}\,S^{\alpha
\beta }({\bf k})e^{i{\bf k\cdot x}}\;.  \label{spcorr_def}
\end{equation}
To carry out the transform, it is convenient define a ``mass'' $m$ via 
\begin{equation}
4m^2\equiv (1-4\bar{\varrho})\Gamma _{\parallel }\varepsilon ^2  \label{mass}
\end{equation}
To keep the system in the homogeneous phase, it is sufficient to impose $%
m^2>0$. We should note that, in the limit $\varepsilon L_{\Vert }\rightarrow
\infty $, the mean-field phase boundary is given precisely by $m^2=0$ .
Otherwise, for {\em finite }$\varepsilon L_{\Vert }$, the system does not
reach the stability limit until $\bar{\varrho}$ equals $\frac 14\left(
1+\left[ 2\pi /\varepsilon L_{\Vert }\right] ^2\right) $. Further, it is
convenient to rescale the lengths 
\begin{equation}
{\bf x}_{\perp }\rightarrow \frac{{\bf x}_{\perp }}{\Gamma _{\perp }^{\frac 1%
2}}\;,\;\;\;\;\;x_{\parallel }\rightarrow \frac{x_{\parallel }}{\Gamma
_{\parallel }^{\frac 12}}\;,  \label{rescale_length}
\end{equation}
so that ${\bf \Gamma }$ becomes the unit matrix. In terms of these rescaled $%
{\bf x,}$ let us define $r\equiv |{\bf x}|$. Also, we should rescale the
elements of the noise matrix, to keep the notation simple: 
\begin{equation}
\sigma _{\perp }\rightarrow \frac{\sigma _{\perp }}{\Gamma _{\perp }}%
\;,\;\;\;\;\;\sigma _{\parallel }\rightarrow \frac{\sigma _{\parallel }}{%
\Gamma _{\parallel }}\;.  \label{rescale_noise}
\end{equation}
Now, the transform can be carried out exactly. The results are: 
\begin{eqnarray}
G^{++}({\bf x}) &=&\mbox{\boldmath $\nabla\sigma\nabla$}\,\left\{ -A^{++}%
\frac 1{r^{d-2}}+B^{++}\left( \cosh (mx_{\parallel })\left( \frac mr\right)
^{\frac{d-2}2}\mbox{K}_{\frac{d-2}2}(mr)\right) \right\}  \label{G++} \\
G_e^{+-}({\bf x}) &=&\mbox{\boldmath $\nabla\sigma\nabla$}\,\left\{ -A_e^{+-}%
\frac 1{r^{d-2}}+B_e^{+-}\left( \cosh (mx_{\parallel })\left( \frac mr%
\right) ^{\frac{d-2}2}\mbox{K}_{\frac{d-2}2}(mr)\right) \right\}  \label{G1}
\\
G_o^{+-}({\bf x}) &=&\mbox{\boldmath $\nabla\sigma\nabla$}\,\left\{ B_o^{+-}%
\mbox{sgn}(\varepsilon )\sinh (mx_{\parallel })\left( \frac mr\right) ^{%
\frac{d-2}2}\mbox{K}_{\frac{d-2}2}(mr)\right\} \;,  \label{G2}
\end{eqnarray}
where $G_{e,o}^{+-}$ are the parts of $G^{+-}$ even or odd in $x_{\parallel
} $, being the transforms of the real and imaginary parts of $S^{+-}$. The
full correlation is, of course, 
\begin{equation}
G^{+-}({\bf x})=G_e^{+-}({\bf x})+G_o^{+-}({\bf x})\;.  \label{split_G+-}
\end{equation}
This decomposition simply reflects the symmetries of the system in the
presence of the field (\ref{g+-}). In eqns. (\ref{G++}-\ref{G2}), $\mbox{K}%
_\nu (z)$ is the modified Bessel function and, provided $\varepsilon \neq 0$%
, the $A$'s and $B$'s are positive constants depending on $\Gamma _{\perp
,\parallel }$ and $\bar{\varrho}$.

The first terms in (\ref{G++}) and (\ref{G1}), being proportional to 
${r^{-d}}$, are the well known power law decays \cite{powers} due to
``FDT-violation'' \cite{ZHSL}. The $B$-terms in (\ref{G++}-\ref{G2}) produce
an exponential decay for large $r$, {\em except} along the field. This can
be easily deduced, since, for large $\mid x_{\parallel }\mid $, the
exponentials of the hyperbolic and Bessel functions cancel. Without
presenting the detailed asymptotic expansions of (\ref{G++}-\ref{G2}), we
illustrate the main results as $mr\rightarrow \infty $.

For ${\bf x}_{\perp }\neq {\bf 0}$ : 
\begin{equation}
G^{++}({\bf x})\;,\;G^{+-}({\bf x})\propto \frac{\sigma _{\parallel
}-\sigma _{\perp }}{r^d}\left[ \frac{{\bf x}_{\perp }^2-(d-1)x_{\parallel }^2%
}{r^2}\right] +...\;,  \label{gen_power}
\end{equation}
where the $...$ represent exponential, short-ranged tails. In this form, we
emphasize the three key ingredients of the ``FDT-violating'' power law
decays, namely, the dependence on $\sigma _{\parallel }\neq \sigma _{\perp }$
(or $\sigma _{\parallel }/\Gamma _{\parallel }\neq \sigma _{\perp }/\Gamma
_{\perp }$), the dipole amplitude and, of course, the $r^{-d}$. Also, note
that the part of $G^{+-}$ odd in $x_{\parallel }$ is seen to be only
short-ranged.

The more interesting limit is for ${\bf x}_{\perp }={\bf 0}$. Then, in
addition to the above power laws, we have another power: $\mid x_{\parallel
}\mid ^{-(d+1)/2}$. For all $d>1$, this power will dominate over the
``FDT-violating'' component. In fact, if we study $d>3$ systems, even the 
{\em next} leading term in the asymptotic expansion of K$_\nu $ will be more
important than $r^{-d}$. For typical simulations in $d=2$, we would write: 
\begin{eqnarray}
G^{++}({\bf 0},x_{\parallel })\;,\;G_e^{+-}({\bf 0},x_{\parallel })
&\propto &\,-\mid x_{\parallel }\mid ^{-\frac 32}+{\cal O}\left( \mid
x_{\parallel }\mid ^{-2}\right) \;,  \nonumber \\
G_o^{+-}({\bf 0},x_{\parallel }) &\propto &-\mbox{sgn}(\varepsilon
x_{\parallel })\mid x_{\parallel }\mid ^{-\frac 32}\,+{\cal O}\left( \mid
x_{\parallel }\mid ^{-\frac 52}\right) \;.  \label{new power}
\end{eqnarray}
Note that, to emphasize the sign of the amplitude of the leading power, we
included explicit factors of $(-1)$, so that the proportionality constants
in (\ref{new power}) are positive. In conclusion, the spatial correlations
are dominated by the expected $r^{-d}$ power law, except along the field,
where a novel $\mid x_{\parallel }\mid ^{-(d+1)/2}$ decay takes over. This
new power law comes from the coupling between the two species as a result
of the excluded volume constraint and the opposite bias.

Finally, let us turn to comparisons with simulation results. Typically, we
find that power law tails are difficult to observe, in relatively small
systems such as ours. Thus, we focus on the structure factors. Deferring
details to a later publication, we point out some general characteristics.
On the whole, the agreement between data (Fig. \ref{fig1}) and (\ref{exp_str}%
) is quite impressive, considering that the latter is only the ``first
approximation''. For example,

\begin{itemize}
\item  in all three cases, $S(k_{\perp },0)$ is independent of $k_{\bot }$;

\item  the value where $\mbox{Re}\{S^{+-}(0,k_{\Vert })\}$ vanishes agrees
well with the theoretically predicted $k_{\Vert }=\sqrt{(1-3\bar{\rho})/(1-%
\bar{\rho})}\,\varepsilon $;

\item  similarly, the ratio $\mbox{Re}\{S^{+-}(k_{\perp
},0)\}/S^{++}(k_{\perp },0)$ is found to be quite close to the expected
value of $-\bar{\rho}/(1-\bar{\rho})$.
\end{itemize}

On the other hand, the data and (\ref{exp_str}) do not agree so well for,
e.g., $\mbox{Re}\{S^{+-}(0,\frac{2\pi }{L_{\parallel }})\}/S^{++}(0,\frac{%
2\pi }{L_{\parallel }})$. We suspect the origin of this discrepancy to be
the following. Given that our data were collected quite close to a second
order transition (so as to observe relatively large fluctuations), we should
expect the longitudinal component of the two-point function to suffer
considerable renormalization. Certainly, this is the case in the proto model 
\cite{BS_RKPZ,KLS,JS_LC}. It would be very interesting to take into account
finite size corrections, to compute these effects in field theory, and to
carry out extensive runs for detailed comparison.

In summary, we have examined the spatial correlations and the structure
factors in a simple model of biased diffusion of two species. Using both
simulation and analytic techniques, we find the expected power law decay, $%
r^{-d}$, typical of non-equilibrium steady states of a system with
anisotropy and subjected to a conservation law. In addition, a novel power, $%
r^{-(d+1)/2}$, is found for correlations along the bias. The general
agreement between simulations and a simple mean-field theory is surprisingly
good, while we await a better theory in order to make detailed quantitative
comparisons.

\section*{Acknowledgements}

We thank  R. Bausch and Z. Toroczkai for many stimulating discussions. 
This research is supported in part by grants from the National 
Science Foundation through the Division of Materials Research .

\end{document}